\documentclass[11pt,letterpaper]{article}

\usepackage[top=3cm,bottom=3cm,left=3cm,right=3cm]{geometry}
\usepackage[square,comma,numbers,sort&compress]{natbib}
\usepackage{url}
\usepackage{graphicx}
\usepackage{amsmath}

\begin{document}

\title{Measuring the Evolution of Contemporary Western\\ Popular Music}

\author{Joan Serr\`a$^{1}$, \'Alvaro Corral$^{2}$, Mari\'an Bogu\~n\'a$^{3}$, Mart\'in Haro$^{4}$, and Josep Ll.~Arcos$^{1}$\\
\\ 
\normalsize{$^{1}$ Artificial Intelligence Research Institute, Spanish National Research Council (IIIA-CSIC),}\\
\normalsize{Bellaterra, Barcelona, Spain.}\\
\normalsize{$^{2}$ Complex Systems Group, Centre de Recerca Matem\`atica, Bellaterra, Barcelona, Spain.}\\ 
\normalsize{$^{3}$ Departament de F\'isica Fonamental, Universitat de Barcelona, Barcelona, Spain.}\\ 
\normalsize{$^{4}$ Music Technology Group, Universitat Pompeu Fabra, Barcelona, Spain.}\\ 
}

\date{}

\maketitle


\begin{abstract}

Popular music is a key cultural expression that has captured listeners' attention for ages. Many of the structural regularities underlying musical discourse are yet to be discovered and, accordingly, their historical evolution remains formally unknown. Here we unveil a number of patterns and metrics characterizing the generic usage of primary musical facets such as pitch, timbre, and loudness in contemporary western popular music. Many of these patterns and metrics have been consistently stable for a period of more than fifty years, thus pointing towards a great degree of conventionalism. Nonetheless, we prove important changes or trends related to the restriction of pitch transitions, the homogenization of the timbral palette, and the growing loudness levels. This suggests that our perception of the new would be rooted on these changing characteristics. Hence, an old tune could perfectly sound novel and fashionable, provided that it consisted of common harmonic progressions, changed the instrumentation, and increased the average loudness.

\end{abstract}



\section*{Introduction}

Isn't it always the same? This question could be easily posed while listening to the music of any mainstream radio station in a western country. Like language, music is a human universal involving perceptually discrete elements displaying organization~\cite{Patel07BOOK}. Therefore, contemporary popular music may have a well-established set of underlying patterns and regularities~\cite{Patel07BOOK,Ball10BOOK,Huron06BOOK,Honing11BOOK}, some of them potentially inherited from the classical tradition~\cite{Lerdahl83BOOK,Temperley07BOOK}. Yet, as an incomparable artistic product for conveying emotions~\cite{Juslin01BOOK}, music must incorporate variation over such patterns in order to play upon people's memories and expectations, making it attractive to listeners~\cite{Huron06BOOK,Honing11BOOK}. For the very same reasons, long-term variations of the underlying patterns may also occur across years~\cite{Reynolds05NATURE}. Many of these aspects remain formally unknown or lack scientific evidence, specially the latter, which is very often neglected in music-related studies, from musicological analyses to technological applications. The study of patterns and long-term variations in popular music could shed new light on relevant issues concerning its organization, structure, and dynamics~\cite{Zanette08NATURE}. More importantly, it addresses valuable questions for the basic understanding of music as one of the main expressions of contemporary culture: Can we identify some of the patterns behind music creation? Do musicians change them over the years? What makes a popular tune novel to us? Is there an `evolution' of musical discourse?

Current technologies for music information processing~\cite{Casey08IEEE,Muller11JSTSP} provide a unique opportunity to answer the above questions under objective, empirical, and quantitative premises. Moreover, akin to recent advances in other cultural assets~\cite{Michel11SCIENCE}, they allow for unprecedented large-scale analyses. One of the first publicly-available large-scale collections that has been analyzed by standard music processing technologies is the million song dataset~\cite{BertinMahieux11ISMIR}. Among others, the dataset includes the year annotations and audio descriptions of 464,411 distinct music recordings (from 1955 to 2010), which roughly corresponds to more than 1,200 days of continuous listening. Such recordings span a variety of popular genres, including rock, pop, hip hop, metal, or electronic. Explicit descriptions available in the dataset~\cite{Jehan05THESIS} cover three primary and complementary musical facets~\cite{Ball10BOOK}: loudness, pitch, and timbre. Loudness basically correlates with our perception of sound amplitude or volume\footnote{Notice that we consider the intrinsic loudness of a recording, not the loudness one could manipulate by e.g.~changing the volume of his/her preferred music-listening device.}. Pitch roughly corresponds to the harmonic content of the piece, including its chords, melody, and tonal arrangements. Timbre accounts for the sound color, texture, or tone quality, and can be essentially associated with instrument types, recording techniques, and some expressive performance resources. These three music descriptions can be obtained at the temporal resolution of the beat, which is perhaps the most relevant temporal unit in music, specially in western popular music~\cite{Ball10BOOK,Honing11BOOK}.

Here we study the music evolution under the aforementioned premises and large-scale resources. By exploiting tools and concepts from statistical physics and complex networks~\cite{Bak96BOOK,Newman05CP,Newman10BOOK,Barrat08BOOK}, we unveil a number of statistical patterns and metrics characterizing the general usage of pitch, timbre, and loudness in contemporary western popular music. Many of these patterns and metrics remain consistently stable for a period of more than 50 years, which points towards a great degree of conventionalism in the creation and production of this type of music. Yet, we find three important trends in the evolution of musical discourse: the restriction of pitch sequences (with metrics showing less variety in pitch progressions), the homogenization of the timbral palette (with frequent timbres becoming more frequent), and growing average loudness levels (threatening a dynamic richness that has been conserved until today). This suggests that our perception of the new would be essentially rooted on identifying simpler pitch sequences, novel timbral mixtures, and louder volumes. Hence, an old tune with slightly simpler chord progressions, new instrument sonorities, and recorded with modern techniques that allowed for increased loudness levels could be easily perceived as novel, fashionable, and groundbreaking.


\section*{Results}

\begin{figure}[!tb]
	\begin{center}
	\includegraphics[width=1\linewidth]{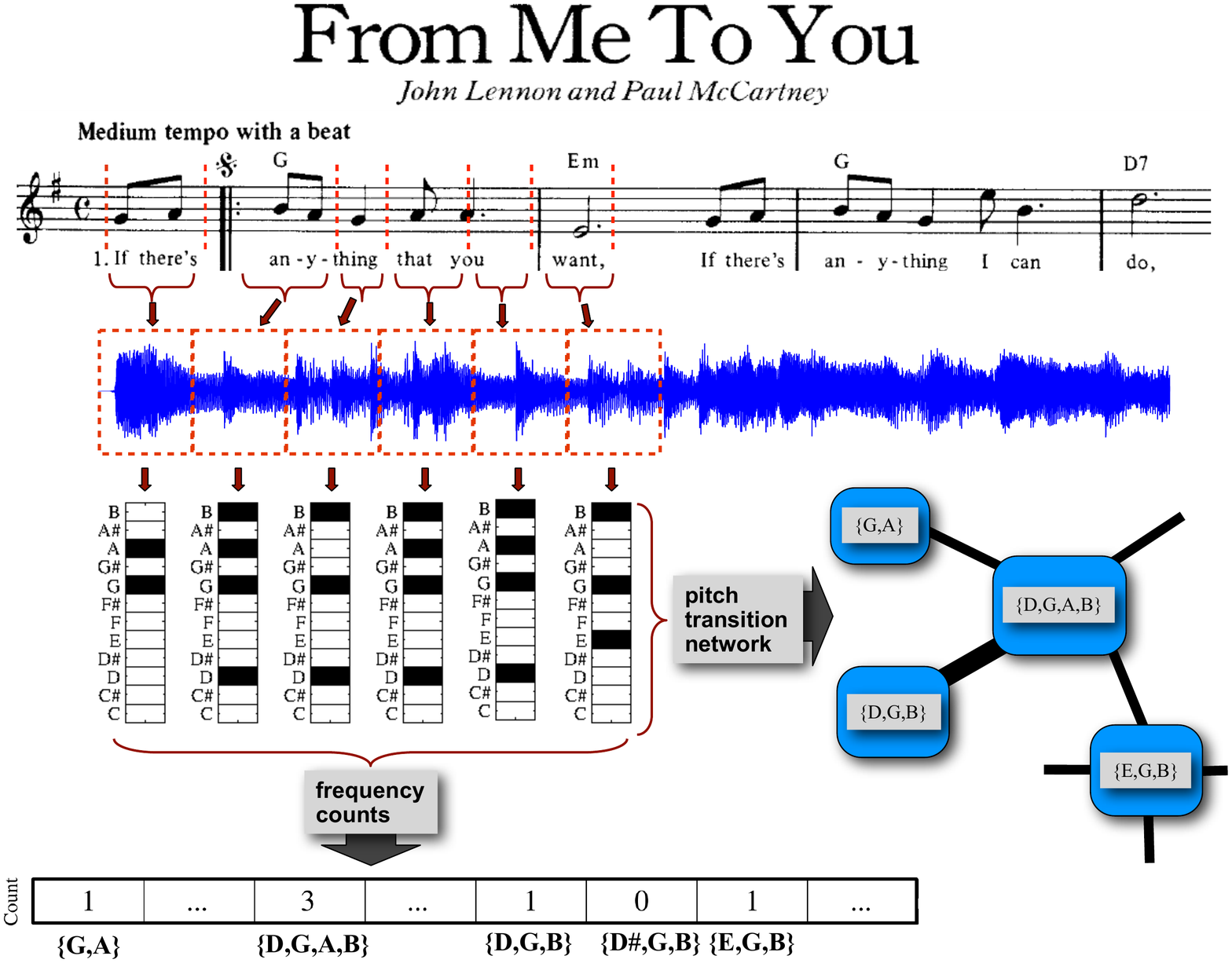}
	\end{center}		
	\caption{Method schematic summary with pitch data. The dataset contains the beat-based music descriptions of the audio rendition of a musical piece or score (G, Em, and D7 on the top of the staff denote chords). For pitch, these descriptions reflect the harmonic content of the piece~\cite{Jehan05THESIS}, and consist of the 12 pitch class relative energies, also called chromas~\cite{Casey08IEEE,Muller11JSTSP} (a pitch class C, C\#, D, etc.\ is the set of all pitches that are a whole number of octaves apart, e.g.~notes C1, C2, and C3 all collapse to pitch class C). Descriptions are encoded into music codewords, using a binary discretization in the case of pitch. Codewords are subsequently used to perform frequency counts and to build a complex network.}
	\label{fig:Summary}
\end{figure}

To identify structural patterns of musical discourse we first need to build a `vocabulary' of musical elements (Fig.~\ref{fig:Summary}). To do so, we encode the dataset descriptions by a  discretization of their values, yielding what we call music \textit{codewords}~\cite{Haro11PLOS} (see Supplementary Materials,~SM). Next, to quantify long-term variations of this vocabulary, we need to obtain samples of it at different periods of time. For that we perform a Monte Carlo sampling of one million beat-consecutive codewords per year period considering entire tracks. We choose a 5-year period and sample 10 times at each possible central year\footnote{For instance, at the central year of 1994 we sample one million consecutive beats by choosing full tracks whose year annotation is between 1992 and 1996, both included.}, guaranteeing a representative sample with a smooth evolution.

\begin{figure*}[!tb]
	\begin{minipage}[!t]{0.49\linewidth}
	\includegraphics{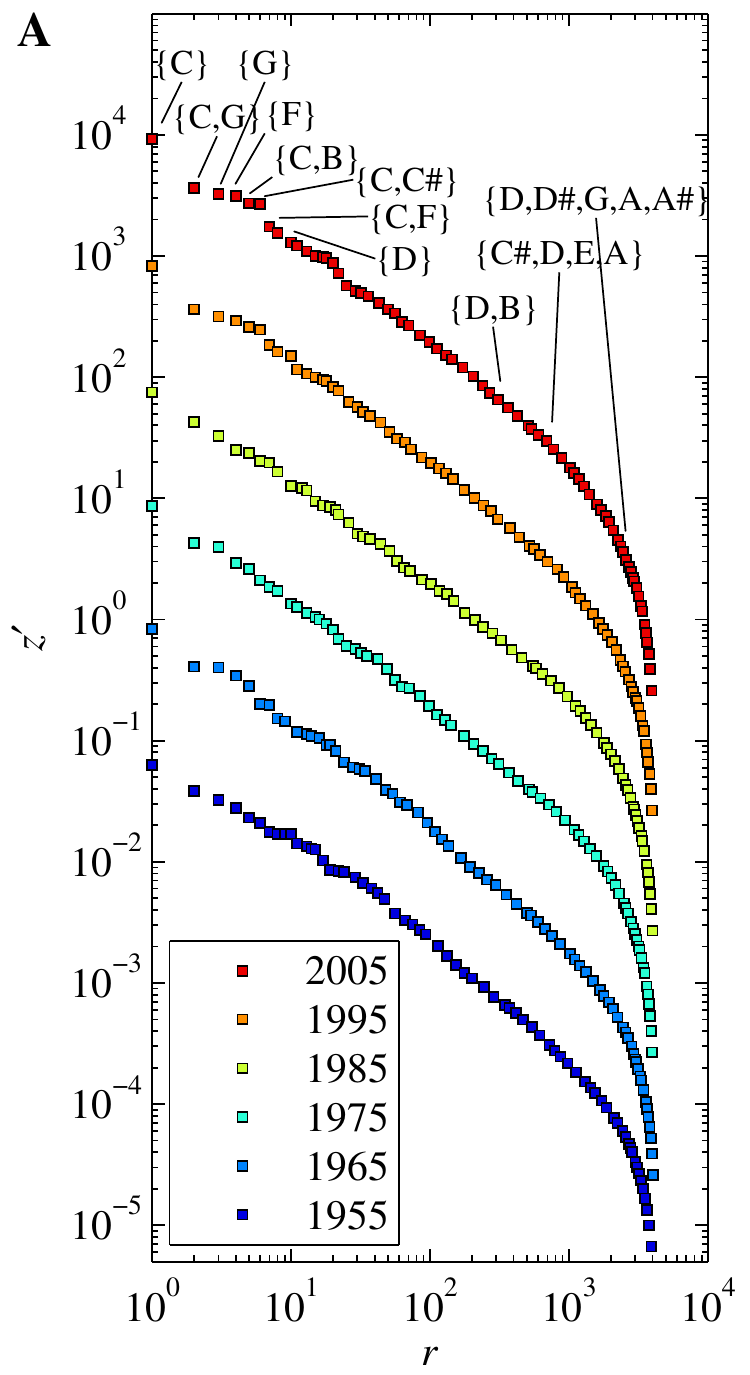}
	\end{minipage}
	\begin{minipage}[!t]{0.49\linewidth}
	\includegraphics{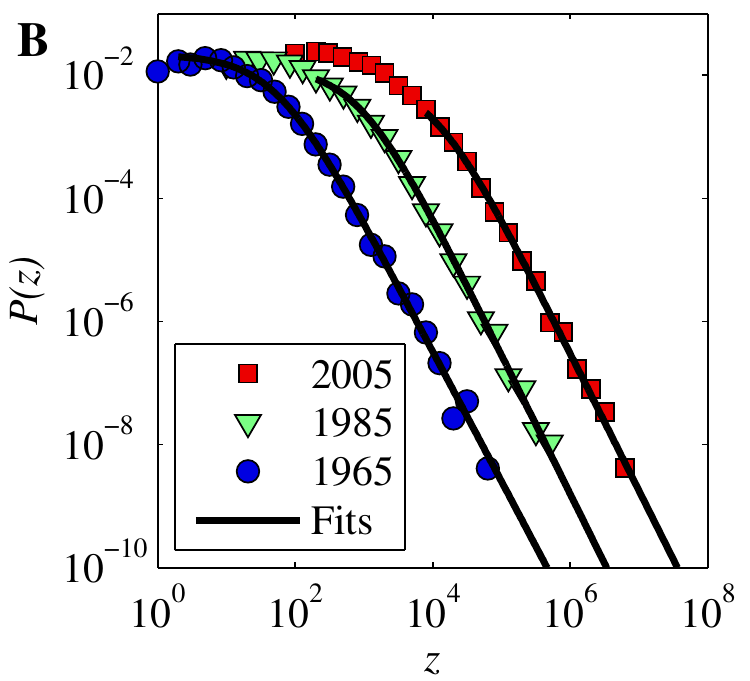} \\
	\includegraphics{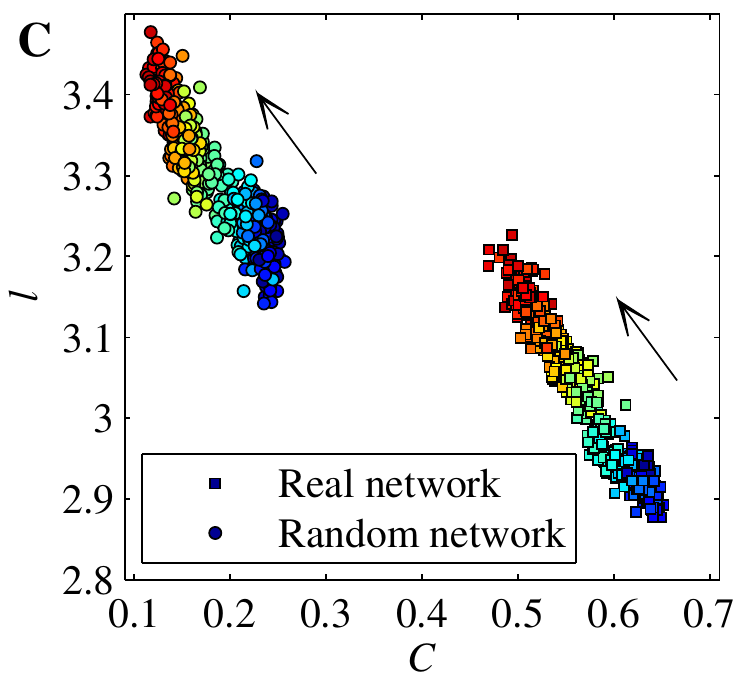}
	\end{minipage}
	\caption{(A) Examples of the rank-frequency distribution for pitch codewords (relative frequencies $z'$ such that $\sum_r z'_r=1$). For ease of visualization, curves are chronologically shifted by a factor of 10 in the vertical axis. Some frequent and infrequent codewords are shown. (B) Examples of the density values and their fits, taking $z$ as the random variable. Curves are chronologically shifted by a factor of 10 in the horizontal axis. (C) Average shortest path length $l$ versus clustering coefficient $C$ for pitch networks (right) and their randomized versions (left). Values calculated without considering the 10 highest degree nodes (see~SM). Arrows indicate chronology (red and blue colors indicate values for more and less recent years, respectively).} 
	\label{fig:Pitch}
\end{figure*}

We first count the frequency of usage of pitch codewords (i.e.~the number of times each codeword type appears in a sample). We observe that most used pitch codewords generally correspond to well-known harmonic items~\cite{DeClercq11PM}, while unused codewords correspond to strange/dissonant pitch combinations (Fig.~\ref{fig:Pitch}A). Sorting the frequency counts in decreasing order provides a very clear pattern behind the data: a power law~\cite{Newman05CP} of the form $z \propto r^{-\alpha}$, where $z$ corresponds to the frequency count of a codeword, $r$ denotes its rank (i.e.~$r=1$ for the most used codeword and so forth), and $\alpha$ is the power law exponent. Specifically, we find that the distribution of codeword frequencies for a given year nicely fits to $P(z) \propto (c+z)^{-\beta}$ for $z>z_{\text{min}}$, where we take $z$ as the random variable~\cite{Adamic02G}, $\beta=1+1/\alpha$ as the exponent, and $c$ as a constant (Fig.~\ref{fig:Pitch}B). A power law indicates that a few codewords are very frequent while the majority are highly infrequent (intuitively, the latter provide the small musical nuances necessary to make a discourse attractive to listeners~\cite{Huron06BOOK,Honing11BOOK}). Nonetheless, it also states that there is no characteristic frequency nor rank separating most used codewords from largely unused ones (except for the largest rank values due to the finiteness of the vocabulary). Another non-trivial consequence of power-law behavior is that extreme events (i.e.~very rare codewords) will certainly show up in a discourse providing the listening time is sufficient.

Importantly, we find this power-law behavior to be invariant across years, with practically the same fit parameters. In particular, the exponent $\beta$ remains close to an average of $2.18\pm 0.06$, which is similar to Zipf's law in linguistic text corpora~\cite{Zipf49BOOK} and contrasts with the exponents found in previous small-scale, symbolic-based music studies~\cite{Zanette06MS,BeltrandelRio08PA}. The slope of the least squares linear regression of $\beta$ as a function of the year is negligible within statistical significance ($p>0.05$, t-test). This makes a high stability of the distribution of pitch codeword frequencies across more than 50 years of music evident. However, it could well be that, even though the distribution is the same for all years, codeword rankings were changing (e.g.~a certain codeword was used frequently in 1963 but became mostly unused by 2005). To assess this possibility we compute the Spearman's rank correlation coefficients~\cite{Hollander99BOOK} for all possible year pairs and find that they are all extremely high, with an average of $0.97\pm 0.02$ and a minimum above 0.91. These high correlations indicate that codeword rankings practically do not vary with years.

Codeword frequency distributions provide a generic picture of vocabulary usage. However, they do not account for discourse syntax, as well as a simple selection of words does not necessarily constitute an intelligible sentence. One way to account for syntax is to look at local interactions or transitions between codewords, which define explicit relations that capture most of the underlying regularities of the discourse and that can be directly mapped into a network or graph~\cite{Newman10BOOK,Barrat08BOOK}. Hence, analogously to language-based analyses~\cite{Sigman02PNAS,FerreriCancho04PRE,Amancio11NJP}, we consider the transition networks formed by codeword successions, where each node represents a codeword and each link represents a transition (see SM). The topology of these networks and common metrics extracted from them can provide us with valuable clues about the evolution of musical discourse.

All the transition networks we obtain are sparse, meaning that the number of links connecting codewords is of the same order of magnitude as the number of codewords. Thus, in general, only a limited number of transitions between codewords is possible. Such constraints would allow for music recognition and enjoyment, since these capacities are grounded in our ability for guessing/learning transitions~\cite{Huron06BOOK,Honing11BOOK,Juslin01BOOK} and a non-sparse network would increase the number of possibilities in a way that guessing/learning would become unfeasible. Thinking in terms of originality and creativity, a sparse network means that there are still many `composition paths' to be discovered. However, some of these paths could run into the aforementioned guessing/learning tradeoff~\cite{Reynolds05NATURE}. Overall, network sparseness provides a quantitative account of music's delicate balance between predictability and surprise.

In sparse networks, the most fundamental characteristic of a codeword is its degree $k$, which measures the number of links to other codewords. With pitch networks, this quantity is distributed according to a power law $P(k) \propto k^{-\gamma}$ for $k>k_{\text{min}}$, with the same fit parameters for all considered years. The exponent $\gamma$, which has an average of $2.20\pm 0.06$, is similar to many other real complex networks~\cite{Newman10BOOK}, and the median of the degree $k$ is always 4. Nevertheless, we observe important trends in the other considered network metrics, namely the average shortest path length $l$, the clustering coefficient $C$, and the assortativity with respect to random $\Gamma$. Specifically, $l$ slightly increases from 2.9 to 3.2, values comparable to the ones obtained when randomizing the network links. The values of $C$ show a considerable decrease from 0.65 to 0.45, and are much higher than those obtained for the randomized network. Thus, the small-worldness~\cite{Watts98NATURE} of the networks decreases with years (Fig.~\ref{fig:Pitch}C). This trend implies that the reachability of a pitch codeword becomes more difficult. The number of hops or steps to jump from one codeword to the other (as reflected by $l$) tends to increase and, at the same time, the local connectivity of the network (as reflected by $C$) tends to decrease. Additionally, $\Gamma$ is always below 1, which indicates that the networks are always less assortative than random (i.e.~well-connected nodes are less likely to be connected among them), a tendency that grows with time. A joint reduction of the small-worldness and the network assortativity stresses a progressive restriction of pitch transitions, with less transition options and more defined paths between codewords (intuitively, the most conventional ones).

\begin{figure*}[!tb]
	\begin{center}
	\includegraphics{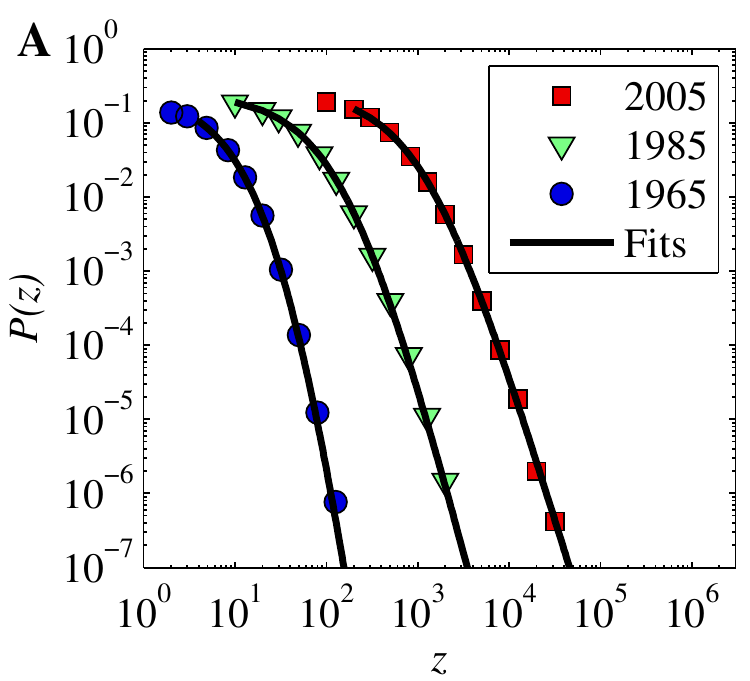}
	\includegraphics{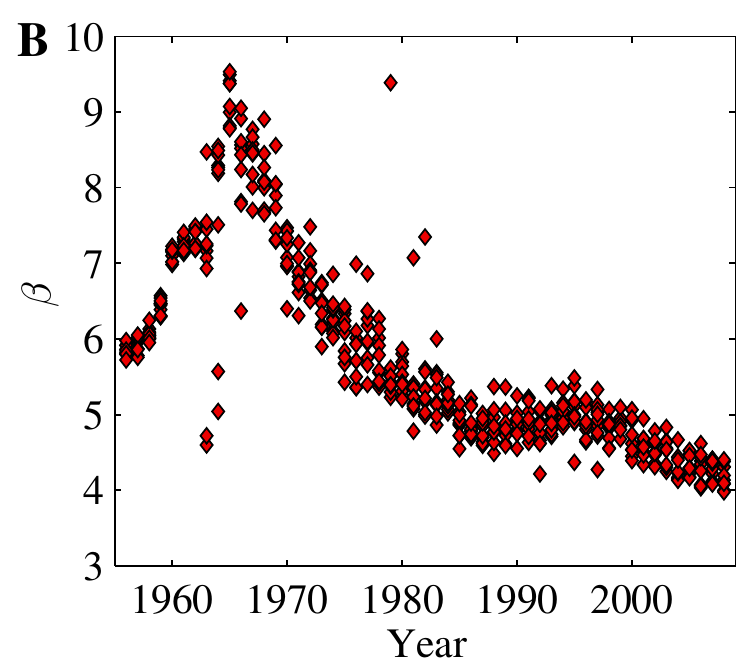} \\ ~\vspace{-0.1cm} \\
	\includegraphics{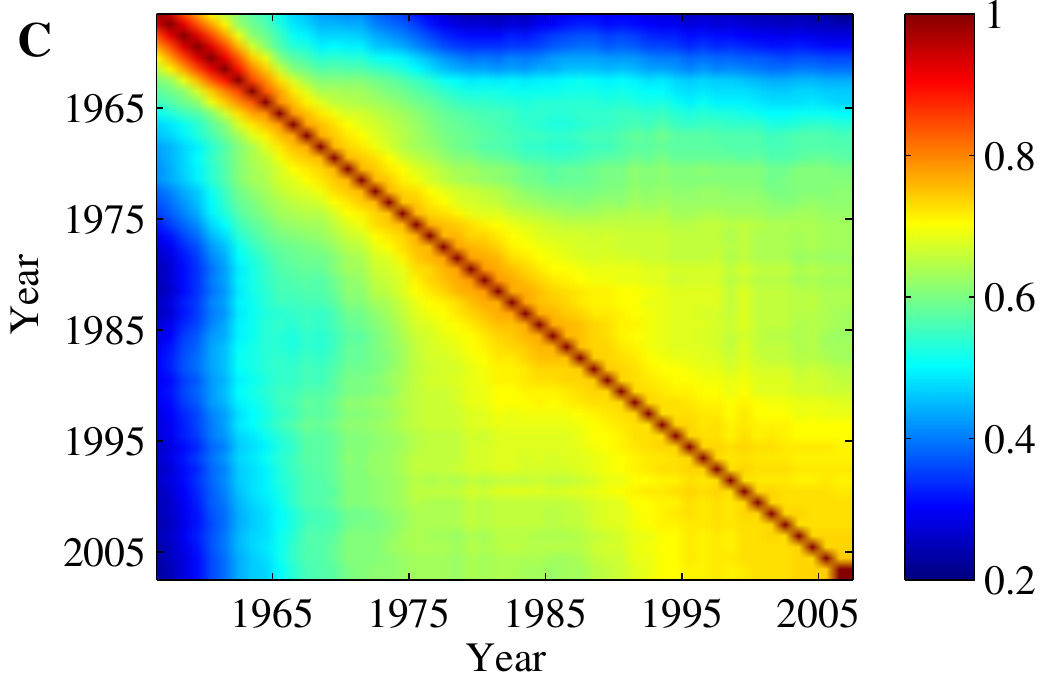}
	\end{center}
	\vspace{-0.7cm}
	\caption{(A) Examples of the density values and fits of timbre codewords taking $z$ as the random variable. (B) Fitted exponents $\beta$. (C) Spearman's rank correlation coefficients for all possible year pairs.}
	\label{fig:Timbre}
\end{figure*}

As opposed to pitch, timbre provides a different picture. Even though the distribution of timbre codeword frequencies is also well-fitted by a power law (Fig.~\ref{fig:Timbre}A), the parameters of this distribution vary across years. In particular, since 1965, $\beta$ constantly decreases to values approaching 4 (Fig.~\ref{fig:Timbre}B). Smaller values of $\beta$ indicate less timbral variety: frequent codewords become more frequent, and infrequent ones become even less frequent. This evidences a growing homogenization of the global timbral palette. Interestingly, rank correlation coefficients are generally below 0.7, with an average of $0.57\pm 0.15$ (Fig.~\ref{fig:Timbre}C). These rather low rank correlations would act as an attenuator of the sensation that contemporary popular music is becoming more homogeneous, timbrically speaking. The fact that frequent timbres of a certain time period become infrequent after some years could mask global homogeneity trends to listeners.

Compared to timbre codeword frequencies, metrics obtained from timbre transition networks show no substantial variation. Again, similar median degrees (all equal to 8) and degree distributions were observed for all considered years. However, we were not able to achieve a proper fit for the latter (SM). The values of $l$ fluctuate around 4.8 and $C$ is always below 0.01. Noticeably, both are close to the values obtained with randomly wired networks. Values of $\Gamma$ are larger than 1, constantly increasing since 1965. Thus, in contrast to pitch, timbre networks are more assortative than random.

\begin{figure*}[!tb]
	\includegraphics{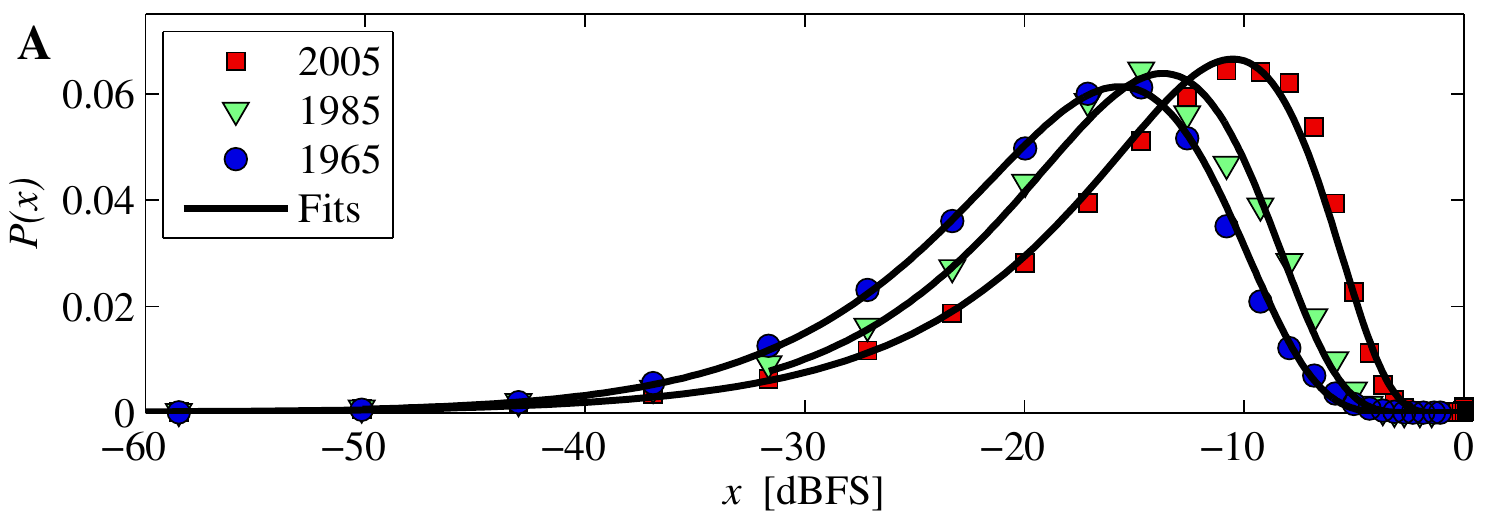} \\ ~\vspace{-0.1cm} \\
	\includegraphics{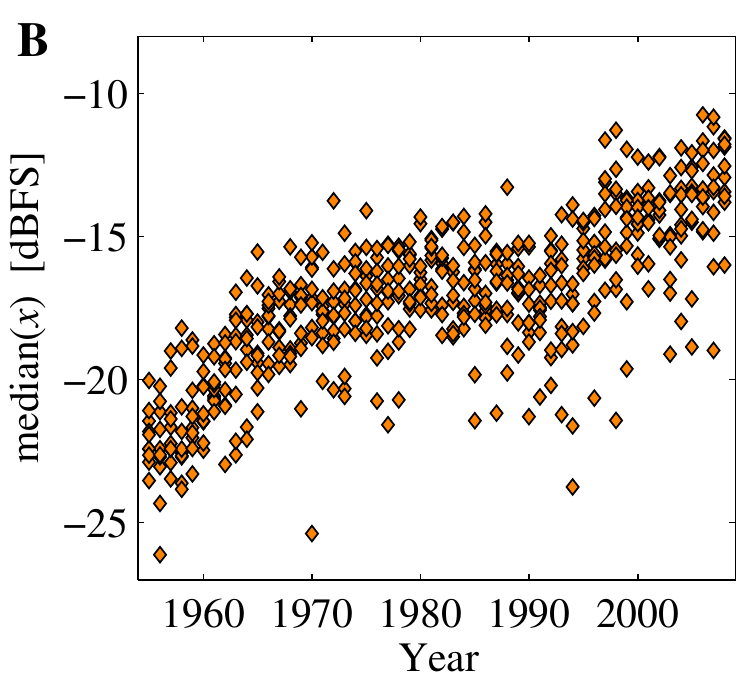}
	\includegraphics{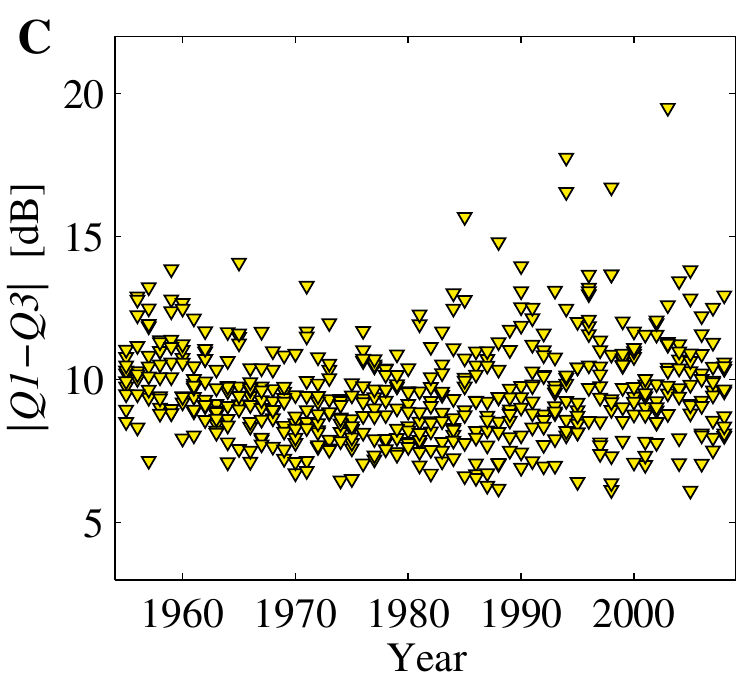} 
	\caption{(A) Examples of the density values and fits of the loudness variable $x$. (B) Empiric distribution medians. (C) Dynamic variation, expressed as absolute differences between the first and third quartiles of $x$, $|Q_1-Q_3|$.}
	\label{fig:Loudness}
\end{figure*}

Loudness distributions are generally well-fitted by a reversed log-normal function (Fig.~\ref{fig:Loudness}A). Plotting them provides a visual account of the so-called loudness race (or loudness war), a terminology that is used to describe the apparent competition to release recordings with increasing loudness~\cite{Milner09BOOK,Deruty11SOS}, perhaps with the aim of catching potential customers' attention in a music broadcast. The empiric median of the loudness values $x$ grows from $-22$~dB$_{\text{FS}}$ to $-13$~dB$_{\text{FS}}$ (Fig.~\ref{fig:Loudness}B), with a least squares linear regression yielding a slope of 0.13~dB/year ($p<0.01$, t-test). In contrast, the absolute difference between the first and third quartiles of $x$ remains constant around 9.5~dB (Fig.~\ref{fig:Loudness}C), with a regression slope that is not statistically significant ($p>0.05$, t-test). This shows that, although music recordings become louder, their absolute dynamic variability has been conserved. However, and perhaps most importantly, one should notice that digital media cannot output signals over 0~dB$_{\text{FS}}$~\cite{Oppenheim99BOOK}, which severely restricts the possibilities for maintaining the dynamic variability if the median continues to grow. 

Finally we look at loudness transition networks, which show comparable degree distributions, a median degree between 13 and 14, values of $l$ between 8 and 10, and a $\Gamma$ fluctuating around 1.08. Noticeably, $l$ is appreciably beyond the values obtained by randomly wired networks. The values of $C$ have an average of $0.59\pm 0.02$, also much above the values obtained by the random networks. These two observations suggest that the network has a one-dimensional character, inferring that no extreme loudness transitions occur. The very stable metrics obtained for loudness networks imply that, despite the race towards louder music, the topology of loudness transitions is maintained.


\section*{Discussion}

Much of the gathered evidence points towards an important degree of conventionalism, in the sense of blockage or no-evolution, in the creation and production of contemporary western popular music. Thus, from a global perspective, popular music would have no clear trends and show no considerable changes  in more than fifty years. Pitch codeword frequencies are found to be always under the same underlying pattern: a power law with the same exponent and fitting parameters. Moreover, frequency-based rankings of pitch codewords are practically identical, and several of the network metrics for pitch, timbre, and loudness remain immutable. Frequency distributions for timbre and loudness also fall under a universal pattern: a power law and a reversed log-normal distribution, respectively. However, these distributions' parameters do substantially change with years. In addition, some metrics for pitch networks clearly show a progression. Thus, beyond the global perspective, we observe a number of trends in the evolution of contemporary popular music. These point towards less variety in pitch transitions, towards a consistent homogenization of the timbral palette, and towards louder and, in the end, potentially poorer volume dynamics.

Each of us has a perception of what is new and what is not in popular music. According to our findings, this perception should be largely rooted on the simplicity of pitch sequences, the usage of relatively novel timbral mixtures, and the exploitation of modern recording techniques that allow for louder volumes. This brings us to conjecture that an old popular music piece would be perceived as novel by essentially following these guidelines. In fact, it is informally known that a `safe' way for contemporizing popular music tracks is to record a new version of an existing piece with current means, but without altering the main `semantics' of the discourse.

Some of the conclusions reported here have historically remained as conjectures, based on restricted resources, or rather framed under subjective, qualitative, and non-systematic premises. With the present work, we gain empirical evidence through a formal, quantitative, and systematic analysis of a large-scale music collection. We encourage the development of further historical databases to be able to quantify the major transitions in the history of music, and to start looking at more subtle evolving characteristics of particular genres or artists, without forgetting the whole wealth of cultures and music styles present in the world.


\section*{Supplementary Materials}

Supplementary materials including materials, methods, and figures are in a separate file.


\section*{Acknowledgments}

We thank the million song dataset team to make this massive source of data publicly available\footnote{\url{http://labrosa.ee.columbia.edu/millionsong}}. This work was supported by Catalan Government grants 2009-SGR-164 (A.C.), 2009-SGR-1434 (J.S. and J.Ll.A.), 2009-SGR-838 (M.B.) and ICREA Academia Prize 2010 (M.B.), European Comission grant FP7-ICT-2011.1.5-287711 (M.H.), Spanish Government grants FIS2009-09508 (A.C.), FIS2010-21781-C02-02 (M.B.), and TIN2009-13692-C03-01 (J.Ll.A.), and Spanish National Research Council grant JAEDOC069/2010 (J.S.).




\begin{thebibliography}{32}
\providecommand{\natexlab}[1]{#1}
\providecommand{\url}[1]{\texttt{#1}}
\expandafter\ifx\csname urlstyle\endcsname\relax
  \providecommand{\doi}[1]{doi: #1}\else
  \providecommand{\doi}{doi: \begingroup \urlstyle{rm}\Url}\fi

\bibitem[Patel(2007)]{Patel07BOOK}
A.~D. Patel.
\newblock \emph{{Music, language, and the brain}}.
\newblock Oxford University Press, Oxford, UK, 2007.

\bibitem[Ball(2010)]{Ball10BOOK}
P.~Ball.
\newblock \emph{{The music instinct: how music works and why we can't do
  without it}}.
\newblock Bodley Head, London, UK, 2010.

\bibitem[Huron(2006)]{Huron06BOOK}
D.~Huron.
\newblock \emph{{Sweet anticipation: music and the psychology of expectation}}.
\newblock MIT Press, Cambridge, USA, 2006.

\bibitem[Honing(2011)]{Honing11BOOK}
H.~Honing.
\newblock \emph{{Musical cognition: a science of listening}}.
\newblock Transaction Publishers, Piscataway, USA, 2011.

\bibitem[Lerdahl and Jackendoff(1983)]{Lerdahl83BOOK}
F.~Lerdahl and R.~Jackendoff.
\newblock \emph{{A generative theory of tonal music}}.
\newblock MIT Press, Cambridge, USA, 1983.

\bibitem[Temperley(2007)]{Temperley07BOOK}
D.~Temperley.
\newblock \emph{{Music and probability}}.
\newblock MIT Press, Cambridge, USA, 2007.

\bibitem[Juslin and Sloboda(2001)]{Juslin01BOOK}
P.~Juslin and J.~A. Sloboda.
\newblock \emph{{Music and emotion: theory and research}}.
\newblock Oxford University Press, Oxford, UK, 2001.

\bibitem[Reynolds(2005)]{Reynolds05NATURE}
R.~Reynolds.
\newblock {The evolution of sensibility}.
\newblock \emph{Nature}, 434\penalty0 (7031):\penalty0 316--319, 2005.

\bibitem[Zanette(2008)]{Zanette08NATURE}
D.~H. Zanette.
\newblock {Playing by numbers}.
\newblock \emph{Nature}, 453\penalty0 (7198):\penalty0 988--989, 2008.

\bibitem[Casey et~al.(2008)Casey, Veltkamp, Goto, Leman, Rhodes, and
  Slaney]{Casey08IEEE}
M.~A. Casey, R.~Veltkamp, M.~Goto, M.~Leman, C.~Rhodes, and M.~Slaney.
\newblock {Content-based music information retrieval: current directions and
  future challenges}.
\newblock \emph{Proceedings of the IEEE}, 96\penalty0 (4):\penalty0 668--696,
  2008.

\bibitem[M\"{u}ller et~al.(2011)M\"{u}ller, Ellis, Klapuri, and
  Richard]{Muller11JSTSP}
M.~M\"{u}ller, D.~P.~W. Ellis, A.~Klapuri, and G.~Richard.
\newblock {Signal processing for music analysis}.
\newblock \emph{IEEE Journal of Selected Topics in Signal Processing},
  5\penalty0 (6):\penalty0 1088--1110, 2011.

\bibitem[Michel et~al.(2011)Michel, Shen, Aiden, Veres, Gray, Pickett, Hoiberg,
  Clancy, Norvig, Orwant, Pinker, Nowak, and Aiden]{Michel11SCIENCE}
J.-B. Michel, Y.~K. Shen, A.~P. Aiden, A.~Veres, M.~K. Gray, J.~P. Pickett,
  D.~Hoiberg, D.~Clancy, P.~Norvig, J.~Orwant, S.~Pinker, M.~A. Nowak, and
  E.~L. Aiden.
\newblock {Quantitative analysis of culture using millions of digitized books}.
\newblock \emph{Science}, 331\penalty0 (6014):\penalty0 176--182, 2011.

\bibitem[Bertin-Mahieux et~al.(2011)Bertin-Mahieux, Ellis, Whitman, and
  Lamere]{BertinMahieux11ISMIR}
T.~Bertin-Mahieux, D.~P.~W. Ellis, B.~Whitman, and P.~Lamere.
\newblock {The million song dataset}.
\newblock In \emph{Proc. of the Int. Soc. for Music Information Retrieval Conf.
  (ISMIR)}, pages 591--596, 2011.

\bibitem[Jehan(2005)]{Jehan05THESIS}
T.~Jehan.
\newblock \emph{{Creating music by listening}}.
\newblock PhD thesis, Massachussets Institute of Technology, Cambridge, USA,
  2005.

\bibitem[Bak(1996)]{Bak96BOOK}
P.~Bak.
\newblock \emph{{How nature works: the science of self-organized criticality}}.
\newblock Copernicus, New York, USA, 1996.

\bibitem[Newman(2005)]{Newman05CP}
M.~E.~J. Newman.
\newblock {Power laws, Pareto distributions and Zipf's law}.
\newblock \emph{Contemporary Physics}, 46:\penalty0 323--351, 2005.

\bibitem[Newman(2010)]{Newman10BOOK}
M.~E.~J. Newman.
\newblock \emph{{Networks: an introduction}}.
\newblock Oxford University Press, Oxford, UK, 2010.

\bibitem[Barrat et~al.(2008)Barrat, Barth\'{e}lemy, and
  Vespignani]{Barrat08BOOK}
A.~Barrat, M.~Barth\'{e}lemy, and A.~Vespignani.
\newblock \emph{{Dynamical processes on complex networks}}.
\newblock Cambridge University Press, Cambridge, UK, 2008.

\bibitem[Haro et~al.(2012)Haro, Serr\`{a}, Herrera, and Corral]{Haro11PLOS}
M.~Haro, J.~Serr\`{a}, P.~Herrera, and \'{A}. Corral.
\newblock {Zipf's law in short-time timbral codings of speech, music, and
  environmental sound signals}.
\newblock \emph{PLoS ONE}, 7\penalty0 (3):\penalty0 e33993, 2012.

\bibitem[{De Clercq} and Temperley(2011)]{DeClercq11PM}
T.~{De Clercq} and D.~Temperley.
\newblock {A corpus analysis of rock harmony}.
\newblock \emph{Popular Music}, 30\penalty0 (1):\penalty0 47--70, 2011.

\bibitem[Adamic and Huberman(2002)]{Adamic02G}
L.~A. Adamic and B.~A. Huberman.
\newblock {Zipf's law and the internet}.
\newblock \emph{Glottometrics}, 3:\penalty0 143--150, 2002.

\bibitem[Zipf(1949)]{Zipf49BOOK}
G.~K. Zipf.
\newblock \emph{{Human behavior and the principle of least effort}}.
\newblock Addison-Wesley, Boston, USA, 1949.

\bibitem[Zanette(2006)]{Zanette06MS}
D.~H. Zanette.
\newblock {Zipf's law and the creation of musical context}.
\newblock \emph{Musicae Scientiae}, 10\penalty0 (1):\penalty0 3--18, 2006.

\bibitem[{Beltr\'{a}n del R\'{\i}o} et~al.(2008){Beltr\'{a}n del R\'{\i}o},
  Cocho, and Naumis]{BeltrandelRio08PA}
M.~{Beltr\'{a}n del R\'{\i}o}, G.~Cocho, and G.~G. Naumis.
\newblock {Universality in the tail of musical note rank distribution}.
\newblock \emph{Physica A}, 387\penalty0 (22):\penalty0 5552--5560, 2008.

\bibitem[Hollander and Wolfe(1999)]{Hollander99BOOK}
M.~Hollander and D.~A. Wolfe.
\newblock \emph{{Nonparametric statistical methods}}.
\newblock Wiley, New York, USA, 2nd edition, 1999.

\bibitem[Sigman and Cecchi(2002)]{Sigman02PNAS}
M.~Sigman and G.~A. Cecchi.
\newblock {Global organization of the Wordnet lexicon.}
\newblock \emph{Proceedings of the National Academy of Sciences of the USA},
  99\penalty0 (3):\penalty0 1742--1747, 2002.

\bibitem[{Ferrer i Cancho} et~al.(2004){Ferrer i Cancho}, Sol\'{e}, and
  K\"{o}hler]{FerreriCancho04PRE}
R.~{Ferrer i Cancho}, R.~V. Sol\'{e}, and R.~K\"{o}hler.
\newblock {Patterns in syntactic dependency networks}.
\newblock \emph{Physical Review E}, 69:\penalty0 051915, May 2004.

\bibitem[Amancio et~al.(2011)Amancio, Altmann, Oliveira, and
  Costa]{Amancio11NJP}
D.~R. Amancio, E.~G. Altmann, O.~N. Oliveira, and L.~da~F. Costa.
\newblock {Comparing intermittency and network measurements of words and their
  dependence on authorship}.
\newblock \emph{New Journal of Physics}, 13:\penalty0 123024, 2011.

\bibitem[Watts and Strogatz(1998)]{Watts98NATURE}
D.~J. Watts and S.~H. Strogatz.
\newblock {Collective dynamics of `small-world' networks}.
\newblock \emph{Nature}, 393\penalty0 (6684):\penalty0 440--442, 1998.

\bibitem[Milner(2009)]{Milner09BOOK}
G.~Milner.
\newblock \emph{{Perfecting sound forever: an aural history of recorded
  music}}.
\newblock Faber and Faber, London, UK, 2009.

\bibitem[Deruty(2011)]{Deruty11SOS}
E.~Deruty.
\newblock {`Dynamic range' and the loudness war}.
\newblock \emph{Sound on Sound -- September 2011}, pages 22--24, 2011.

\bibitem[Oppenheim et~al.(1999)Oppenheim, Schafer, and Buck]{Oppenheim99BOOK}
A.~V. Oppenheim, R.~W. Schafer, and J.~R. Buck.
\newblock \emph{{Discrete-time signal processing}}.
\newblock Prentice-Hall, Upper Saddle River, USA, 2nd edition, 1999.

\end{thebibliography}
\end{document}